\documentclass[12pt]{article}
\usepackage{epsfig}
\newcommand{\bc}{\begin{center}}
\newcommand{\ec}{\end{center}}
\newcommand{\bmi}{\begin{minipage}}
\newcommand{\emi}{\end{minipage}}
\newcommand{\bdi}{\begin{displaymath}}
\newcommand{\edi}{\end{displaymath}}
\newcommand{\bit}{\begin{itemize}}
\newcommand{\eit}{\end{itemize}}

\newcommand{\Ne}{N\'{e}el}

\newcommand{\bi}{\begin{itemize}}
\newcommand{\ei}{\end{itemize}}
\newcommand{\be}{\begin{equation}}
\newcommand{\ee}{\end{equation}}

\newcommand{\bea}{\begin{eqnarray}}
\newcommand{\eea}{\end{eqnarray}}

\newcommand{\bdm}{\begin{displaymath}}
\newcommand{\edm}{\end{displaymath}}
\newcommand{\beas}{\begin{eqnarray}}
\newcommand{\eeas}{\end{eqnarray}}
\newcommand{\bite}{\begin{itemize}}
\newcommand{\enite}{\end{itemize}}
\newcommand{\ben}{\begin{enumerate}}
\newcommand{\een}{\end{enumerate}}

\newcommand{\kag}{{\it kagom{\'e}{\ }}}

\newcommand{\Sier}{Sierpi{\'n}ski}

\begin{document}

\title{The anisotropic quantum antiferromagnet on the \Sier \
gasket: Ground state and thermodynamics}

\author{
Andreas Voigt \\
The University of Georgia, Center for Simulational Physics \\
Athens, Georgia, 30602, USA \\
and \\
Max-Planck-Institut Magdeburg \\
Sandtorstr.1, 39106 Magdeburg, Germany  \\ 
\vspace{0.3cm} \\
Wolfgang Wenzel  \\
Forschungszentrum Karlsruhe, Institut f{\"u}r Nanotechnologie  \\
Postfach 3640, 76021 Karlsruhe, Germany  \\
\vspace{0.3cm }\\
Johannes Richter  \\
Otto-von-Guericke-Universit\"at Magdeburg  \\
Institut f\"ur Theoretische Physik,  
PF 4120, D-39106 Magdeburg, Germany  \\
\vspace{0.3cm}\\
Piotr Tomczak  \\
Uniwersytet im. Adama Mickiewicza  \\ 
Wydzia{\l} Fizyki, ul.Umultowska 85, 61-614 Pozna{\'n}, Poland
\normalsize{ {\ }} 
}

\maketitle

\normalsize

\begin{abstract}
We investigate an antiferromagnetic s=1/2 quantum spin system with
anisotropic spin exchange on a fractal lattice, the \Sier \ gasket. We
introduce a novel approximative numerical method, the configuration
selective diagonalization (CSD) and apply this method to a the \Sier \
gasket with N=42. Using this and other methods we calculate ground
state energies, spin gap, spin--spin correlations and specific heat
data and conclude that the s=1/2 quantum antiferromagnet on the \Sier \
gasket shows a disordered magnetic ground state with a very short
correlation length of $\xi \approx 1$ and an, albeit very small, spin
gap.  This conclusion holds for Heisenberg as well a for XY exchange.
\end{abstract}

\section{Introduction}
\label{intro}

Low-dimensional quantum antiferromagnets (AFM) have been
intensively investigated since the development of quantum
mechanics in the early twenties \cite{Bethe31}. A renewed interest
was motivated by the discovery of high-temperature
superconductivity~\cite{Bednorz86} and the peculiar interplay of
the magnetic and electronic properties of these systems, where
antiferromagnetism and superconductivity appear in close vicinity
(see e.g. \cite{manousakis91} and ref. therein). Since then
various experimental findings for materials for electronically
one- or twodimensional magnetic systems, such as $CuGeO$ $CaVO$,
$SrCuBO$, have called for more detailed theoretical investigations
of the ground state and the low temperature properties of one- and
two-dimensional quantum AFM (see e.g. \cite{lhuillier01sep} and
ref. therein). Despite many theoretical efforts, many properties
of low-dimensional quantum AFM, in particular the interplay of
quantum fluctuations and magnetic order near quantum critical
points, need further explanation.

The radical difference in the behavior of one- and two-dimensional AFM
has been subject of current debate, in particular with regard to
its interaction with changes in the lattice structure (in the
cuprates), the presence of spin-peierls transitions or the influence of
disorder. One of the most significant dimension-dependent properties is
the type of magnetic order: the ground state of the one-dimensional
s=1/2 Heisenberg chain remains disordered~\cite{Bethe31}, but for the
two--dimensional Heisenberg quantum AFM on square, triangular or
honeycomb lattices one observes a \Ne-like magnetic long range order in
ground state. The dimensional crossover between d=1 and d=2 has been
studied via investigations of ladder structures \cite{dagotto96} or by
varying exchange parameters on two--dimensional lattices (e.g. on the
square lattice or on the triangular lattice)
\cite{affleck94,ihle99,stary02,sindzingre02,nersesyan03,brenig03,sindzingre03}.  

In addition to the dimensionality the spin anisotropy can also
influence the magnetic order in quantum AFM. For example, it is known
that in the one--dimensional linear chain an infinitesimal small
Ising-like exchange aniso\-tropy induces a \Ne \ like magnetic order in
the ground state. In zig-zag ladders the effects of XY anisotropy may
lead to spiral ordering \cite{nersesyan98}. For two--dimensional lattices at
finite temperatures (where the Mermin-Wagner theorem forbids any
\Ne--like long-range order for pure Heisenberg exchange) an XY exchange
anisotropy can induce a vortex type ordering at the Kosterlitz-Thouless
transition \cite{kosterlitz73}.

We have previously studied the influence of dimensionality on the
magnetic order by considering a quantum AFM on a particular lattice
geometry, the \Sier \ gasket, with a topological dimension between one
and two. We considered a Heisenberg interaction between the
nearest-neighbor spins on this lattice and investigated its properties
with exact diagonalization and variational wave functions
\cite{tomczak96co,tomczak96prb}. We supplemented this analysis with
thermodynamical properties using the quantum decimation technique (QDT) 
\cite{tomczak96prb2} and a decoupled-cell Monte-Carlo approach (DCM)
\cite{voigt98jmmm}. Recently we extended this research to higher spins
\cite{voigt01} as well as to anisotropic spin interactions
\cite{voigt02}. From all calculated data we have presented arguments in
favor of a disordered ground state of the \Sier \ gasket quantum AFM.

All mentioned numerical approaches have some limitations and
disadvantages. The exact diagonalization is subject to the well known
constraint on the system size. Its application to quantum spin models
is always challenging but in the case of a fractal lattice even more
complicated due to loss of translational symmetries. Therefore only
small systems with N=15 could be investigated so far (In
\cite{voigt02} we include a system with N=28, which has a similar
topology like N=15 but is in fact not a true \Sier \ gasket). The
variational wave function calculations suffer from the uncertainties in
the reference wavefunction due to lattice frustration and the
thermodynamical properties calculated with QDT and DCM might not probe
the true ground state. Therefore all previous conclusions have been
drawn with particular care and indeed, especially the investigation of
larger systems is very desirable for further support and verification.

In this paper we will apply a new technique, the configuration
selective diagonalization (CSD), in a particular efficient
implementation to investigate larger finite lattices. This approach is
based on ideas developed in quantum chemistry  
\cite{peyer74,stampfuss99} and evolves around an extrapolative
calculation of the ground state and low excitations. Using this
approach we are able to access larger systems in a numerically
controlled approximation. We will present results for the N=42
\Sier \ gasket (the next larger \Sier \ gasket after N=15) for
both the Heisenberg and the XY model. This is to our knowledge the
largest quantum spin system treated so far with a direct
diagonalization technique.  The treatment of this cluster permits the
analysis of spin--spin correlations for larger lattice separations and
allows us to further strengthen the predictions of a disordered ground
state for the Heisenberg model. It also enables us to draw similar
conclusions for the XY model on the \Sier \ gasket.

In section \ref{model} we will introduce the model and its basic
properties. In section \ref{method} we present all numerical
methods used in our investigation. We will emphasize on the CSD in
very detail, as this is the first application of this approach to
quantum spin systems. In section \ref{data} the results of the
calculations will be presented and conclusions about the magnetic
order behavior will be made. We summarize the paper in section
\ref{summ}.

\section{The Model}
\label{model}

\begin{figure}[ht]
\centerline{
\epsfig{file=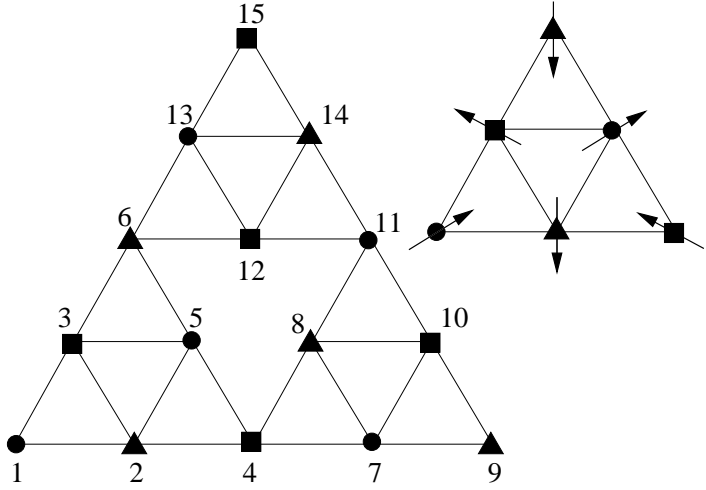,scale=0.50,angle=0}}
\vspace*{0.2cm} \caption{Left: The classical ground state
configuration of the N=15 \Sier \ gasket with 3
sublattices: A (circles), B (squares) and C (triangle). Right: The
spin direction in the classical ground state with an angle of
120$^\circ$ between spins belonging to different sublattices.} 
\label{fig1}
\end{figure}

We consider the quantum s=1/2 AFM with anisotropic spin exchange:

\be \hat H = J \sum_{<i,j>} \left ( {\bf S}^x_i {\bf S}^x_j + {\bf
S}^y_i {\bf S}^y_j + \Delta {\bf S}^z_i {\bf S}^z_j \right ).  \ee
The antiferromagnetic spin exchange $J>0$ is taken between nearest
neighbors on the \Sier \ gasket (an example of this lattice with
N=15 is given in Fig.\ref{fig1}). The anisotropy $\Delta$ will be
studied for the Heisenberg model at $\Delta=1$ and for the XY
model at $\Delta=0$.  The most important geometrical property of
the \Sier \ gasket is its fractal Hausdorff dimension of
$d_f={\ln(3) \over \ln(2)} \approx 1.58$. The number of spins on
this lattice is given by N$={1\over2}(3^n+3)$ with
$n=1,2,3,\ldots$. In the paper we will focus on N=6, 15 and 42
(i.e. n=2,3,4).

The classical ground state is a planar spin state with 3 sublattices
for $\Delta \le 1$. The spins in such a sublattice are
ferromagnetically aligned, between spins belonging to different
sublattices we observe an angle of 120$^\circ$. This ground state is
depicted by the arrows in the right part of Fig.\ref{fig1}.
The classical ground state of the \Sier \ gasket is analogous to the
ground state of the two--dimensional triangular lattice and has no
non-trivial degeneracy.

We would like to mention here that there is an ongoing debate of the
relation between the properties of the classical ground state and the
magnetic order of the quantum ground state. For the \kag lattice it was
argued in \cite{lecheminant97} that the infinite degeneracy of the
classical ground state is closely connected to the high number of
low-lying singlets in the quantum case (and therefore maybe to a
liquid-like quantum ground state.) Ongoing research did show that there
are systems like the planar pyrochlore lattice with a classical
non-trivial degeneracy of the ground state, yet with valence bond type
long range order in the quantum regime \cite{fouet03}. The multi-spin
exchange model is another counter example with {\bf no} non-trivial
classical degeneracy but with a large number of low lying singlets
\cite{liming00}. As it has been discussed already in \cite{voigt01} the
\Sier \ gasket is yet another example as it has {\bf no} non-trivial 
classical degeneracy, even though in the quantum case it also shows a
large number of low-lying singlets and a disordered liquid-like ground
state.

\section{The Methods}
\label{method}

In this paper we will use a variety of methods to investigate the
\Sier \ gasket. For finite quantum AFM of small size the exact
diagonalization (ED) with a Lanczos approach is the tool of choice
to investigate the ground state and the low energy spectrum.
Because of the exponentially growing Hilbert space the approach is
usually limited to systems with up to N=36, only for highly
symmetric lattices  like the square lattice one can reach N=40
\cite{schulenburg03,schulenburg03a}. As already stated the CSD
will be used for the calculation of the ground state and first excited
state of the N=42 \Sier \ gasket. For thermodynamic properties we will
deploy a complete diagonalization (CD) for smaller finite systems and a
quantum decimation technique (QDT).

Whereas the complete diagonalization is routinely used (for more
information see e.g. \cite{numrecip}), the QDT is not known to a wide
audience even though it has been successfully implemented to
investigate low-temperature thermodynamics of different
low--dimensional AFM \cite{tomczak96prb1,tomczak03}. Therefore we will
shortly describe the two basic steps of this approach on the example of
the \Sier \ gasket. In the first step, one splits the Hamiltonian $H$
of an infinite system into Hamiltonians $H_i$ of finite 6-spin
subsystems (see right part of Fig.\ref{fig1}) ($H = \sum_iH_i$).
Recalling that the renormalization group (RG) procedure should preserve
the partition function and symmetry of the system and, additionally,
the decimation procedure should preserve the correlation function, one
traces out some spin degrees of freedom in each finite subsystem. In
the second step one puts these finite subsystems together and obtains a
renormalized Hamiltonian $H^\prime$:
\begin{eqnarray}
\exp\bigg(\sum_iH_i\bigg) \approx \prod_i\exp (H_i) \approx
\nonumber \\ 
\approx \prod_i\exp (H'_i) \approx
\exp\bigg(\sum_i H'_i\bigg) \nonumber.  
\end{eqnarray}
Note that while splitting the Hamiltonian into finite subsystems and
subsequently replacing the true local Hamiltonian by the renormalized
$H^\prime_i$ one neglects the non-com\-mutativity of the spin operators.

This two-step RG transformation enables one to calculate the free energy
per spin as follows:
\begin{equation}
\label{free_en}
-f/k_BT=\sum_{i=0}^\infty 
\bigg(\frac{1}{3}\bigg)^i\,g(K^{(i)}),
\end{equation}
with $K^{(i)}$ representing the $i$-times transformed coupling
constant $K \equiv J/k_BT$ ($k_B$ -- Boltzmann constant and
$T$ -- temperature). The $g$ in the above equation represents the
contribution to the free energy (per spin) from degrees of freedom 
which have been decimated out in one RG transformation. For additional
details of this method see for e.g. \cite{tomczak96prb2,tomczak03}.

As already stated the CSD will be used for the first time for quantum
spin systems and we therefore describe it now in very detail.

\subsection{Configuration Selective Diagonalization}

As noted above exact diagonalization methods are limited to small
systems because of an the exponential increase of the Hilbert space
with the system size.  In certain circumstances, however, the size of
treatable systems may be increased significantly if only approximate
energies and expectation values are required. Here we describe the
adaptation of the CSD for the approximative calculation of low-energy
properties of quantum spin systems in the context of the N=42 \Sier \
gasket. This approach has been developed using the methodology recently
applied in the multi-configuration interaction methods
\cite{stampfuss99,stephan98,stampfuss03} in quantum chemistry. In the
following, we will describe this method and its implementation for the
system at hand, but note that it is applicable to other systems as
well. The division of the whole system into fragments which are exactly
solvable within an exact Lanczos diagonalization is a necessary
precondition for the application of the CSD, as discussed in detail
below.
\begin{figure}[ht]
{\centering \resizebox*{0.4\textwidth}{!}
{\includegraphics{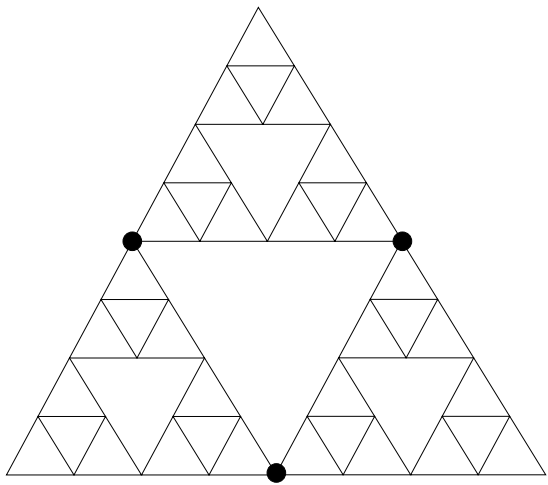}} \par}
\vspace*{0.2cm} \caption{42 Site approximant of the \Sier \
gasket. If the sites indicated by black circles are treated as a
central fragment, the remainder of the system decomposes into
three uncoupled thirteen-site satellite fragments.
\label{fig:g42}}
\end{figure}
The model in Fig.\ref{fig:g42} contains 42 sites, resulting in a
Hilbert space of dimension \( 5.3\times 10^{11} \) in the \(
S_{z}^{tot}=0 \) subspace, which can be reduced by a factor of
approximately four by discrete Abelian symmetries, such as spin- and
real-space reflection. The use of the three-fold rotation incurs too
much computational overhead to be of real value. Even so, it remains
too large for exact diagonalization schemes for low-lying eigenstates.
The special structure of the cluster permits the use of approximate
configuration-selective diagonalization methods. If we subdivide the
cluster into a central fragment containing the three sites indicated
with black dots, its remainder consists of three identical
{}``satellite'' fragments of 13 sites each. If there were no bonds
between the central fragment and the satellites, the eigenstates of
the overall system would be outer products of the eigenstates of the
four fragments. In the presence of interactions, only four bonds
couple the central fragment with each of the satellites, which may be
assumed to perturb the spectrum of the satellites only weakly. Because
the satellite fragments are only weakly coupled to the central system,
it is plausible to expand the eigenstate of the overall system in a
basis of outer products of eigenstates of the satellite fragments. In
the absence of any coupling to the central fragment, this
approximation would be exact, in their presence one can hope that only
a few configuration in this expansion will carry the overwhelming
weight of the wavefunction. This observation suggests the application
of the CSD (which has a long history in similar scenarios in quantum
chemistry \cite{peyer74,shavitt68}) to quantum spin systems.

The basic idea of this approach is simple: Suppose an approximate
wavefunction of the ground state is already known and this
wavefunction has nonzero coefficients only for a small fraction of
the configurations of the Hilbert space. We then estimate the
weight of each remaining unselected configuration in second order
perturbation theory. We keep only such configurations where the
absolute value of the estimated coefficient surpasses a predefined
threshold \( \epsilon \) and sum the perturbative energy
contributions of the neglected configurations. Next we determine
the eigenstate within the new selected Hilbert space with a direct
diagonalization technique. The resulting state will be a better
approximation of the desired eigenstate in the full Hilbert space.
These two steps are alternated with decreasing selection threshold
\( \epsilon \) and the energy  (including the perturbative
correction of the discarded configurations) and other expectation
values are extrapolated to the limit \( \epsilon \rightarrow 0 \).
For many systems this limit can be safely extrapolated with
selected Hilbert spaces that contain only a small fraction of the
possible configurations. The process is initiated with some simple
trial wavefunction containing only the appropriate ground-state
configurations of the segments.

The algorithm thus consists of two distinct phases: in the
\emph{expansion step} new configurations are selected
perturbatively and in the \emph{diagonalization step} the lowest
(or a few of the lowest) eigenvalues of the selected Hilbert space
are determined. In the \emph{diagonalization step} we iteratively
improve a trial vector for the ground state of the selected
Hilbert space using a preconditioned Davidson method
\cite{davidson72,olson90}. The time-consuming step of this
iterative method is the computation of expectation values \(
\left\langle \Psi _{i}|H|\Psi _{j}\right\rangle  \) of the many
body Hamiltonian \( H \) between trial states\begin{equation}
\label{eq:wave} \left| \Psi _{i}\right\rangle =\sum _{k}\alpha
_{ik}\left| \phi _{k}\right\rangle ,
\end{equation}
where \( \left| \phi _{k}\right\rangle  \) designate the configurations
of the selected Hilbert space. The evaluation of such matrix elements
is difficult, because at any given stage, the selected Hilbert space
contains an essentially random subset of the possible configurations.

The Hamilton operator \( H \) of the
system can be written as
\begin{equation}
\label{eq:h}
H=H_{c}+\sum _{s}H_{s}+\sum_s H_{sc}
\end{equation}
where \( s \) enumerates the satellite fragments and \( c \)
designates the central fragment. \( H_{s} \) and \( H_{c} \)
sum terms of H acting on a single fragment, while \( H_{sc} \)
couples the satellite \( s \) to the central fragment.

To evaluate the expectation values we pre-diagonalize the 13-site
satellite fragments in their respective \( S^{tot}_{z} \)
spin-segments and compute the boundary-operators \( S_{sb}^{\pm }
\) and \( S^{z}_{sb} \) for the four boundary sites \( b \) of
each fragment \( s \) in this basis. Similarly we compute the
matrix representation of the corresponding operators for the
central fragment. Each configuration \( \left| \phi
_{k}\right\rangle \)in the Hilbert space is labeled by a
quadruplet of quantum numbers \( (n_{c},n_{1},n_{2},n_{3}) \),
where \( n_{i} \) is the index of an eigenstate of the
corresponding fragment. The first two terms in the Hamiltonian
(\ref{eq:h}) are diagonal in this representation and easily
evaluated.

Nondiagonal terms, in contrast, are difficult to evaluate because of
the sparsity of the selected configurations in the overall Hilbert
space. In order to avoid costly lookup operations we have developed a
so-called residue based scheme for the evaluation of the matrix
elements that we will describe in detail in the following. 
Each coupling term $H_{sc}$ is a sum of products of pairs of
boundary operators described above. To efficiently evaluate the
expectation values of this part of the Hamiltonian we use a
residue based matrix element evaluation technique that was
originally developed for selecting configurations in interaction
methods~\cite{stampfuss99,stephan98}.
\begin{figure}
{\centering \resizebox*{0.6\columnwidth}{!}
{\includegraphics{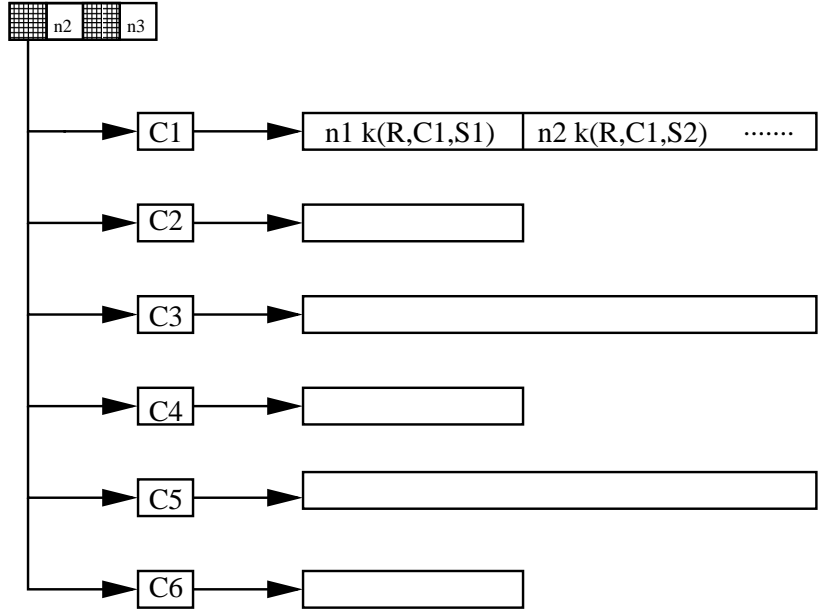}} \par}
\vspace*{0.2cm}
\caption{A single entry \protect\( R\protect \) in the residue tree (top
left) is designated by the quantum numbers of the fragment not changed
by the operator in question (see text, \protect\(
(n_{2},n_{3})\protect \) in the example).  The first level labels the
quantum number of the central fragment that is necessary to completely
specify the state. Attached to each such entry $c_i$ is the list of
indices of the configurations that have quantum numbers
$(c_i,\dots,n_{2},n_{3})$. Each element of such a list contains the
quantum number $n_1$ of the configuration and its associated index 
of the coefficient of the associated configurations \protect\(
k(R,c_{i},s).\protect \)
\label{fig:restree}}
\end{figure}
For a particular configuration of the selected Hilbert space one
individual term in \( H_{cs} \) changes the quantum numbers on the
central fragment and on one of the satellite fragments (in the
following we choose without loss of generality fragment 1). The
quantum numbers on the other two fragments $n_2,n_3$ are the same on
the right- and left-hand side of the configuration for nonzero matrix
elements. All all nonzero contributions arising from this particular
term in the Hamiltonian connect two configurations which have the
same index $n_2,n_3$. Since we wish to enumerate all nonzero matrix
elements, we can use this set as a label for the associated matrix
elements. In the following we will call this set of quantum numbers of
both configurations the transition residue (R) mediating the matrix
element. For each transition residue $(n_2,n_3)$ we can enumerate the
configurations that lead to nonzero matrix elements by a doubly nested
{\em residue tree}. 

All nonzero matrix elements mediated by a particular transition
residue R are uniquely enumerated by pairs of indices of the first
level and pairs of indices of the second level of the residue tree
R. There are no matrix elements spanning different residue trees. As an
example the matrix element evaluation for $H_{sc}=S^+_cS^-_1$ can thus
be written as:
\beas
\langle \Psi _{1}|H_{cs}|\Psi _{2}\rangle & = &
\sum _{R=(n_2,n_3)} \sum_{c_{1},c_{2} {\rm in\ Tree\ R}} \nonumber \\
& & S_{c_{1}c_{2}}^{+}  \large( \sum_{s_{1} {\rm in\ list\ (R,c_1)} }
\sum_{s_{2} {\rm in\ list\ (R,c_2)} } \nonumber \\
& & A^{(1)*}_{(R,c_{1},s_1)} S_{s_{1}s_{2}}^{-} A^{(2)}_{(R,c_{2},s_{2})}
\large) 
\label{eq:mat}
\eeas

In order to carry out the sums in the above equation, we construct a
{\em residue tree} for each $R=(n_2,n_3)$, which is illustrated
schematically in Fig.\ref{fig:restree}. The first level of the tree
enumerates the allowed quantum numbers of the central fragment
$(n_c=c_1,c_2,c_3,\dots)$. The second level of the tree enumerates for
each central fragment quantum number the selected configurations
$(c_i,s_j,n_2,n_3)$ and contains the value of the coefficient of the
associated configuration: $A(R=(n_2,n_3),c_i,s_j,)$. The sum in
Eqn.(\ref{eq:mat}) is then computed by picking all allowed pairs of
branches $c_j,c_j$ of the tree. For each pair of branches, each pair
of entries generates a nonzero matrix element. This double sum can
thus be performed without further lookup operations and evaluates
nevertheless only matrix elements that connect {\em selected
configurations}.  Note that the innermost loops run over all selected
indices on the satellites, i.e. encode O($N_s^2$) operations for a
fully selected Hilbert space, where $N_s \approx 1000$ is the
dimension of the Hilbert space on a satellite fragment. In order to
build the tree just once for many applications of the Hamiltonian, it
is more efficient to store the indices $k(R,c_2,s_2)$ rather than the
coefficients in the tree. 

Using this procedure all matrix elements can be evaluated for
arbitrarily complicated subspaces without lookup operations.  We have
implemented the residue tree by nested Adelson-Vleski-Landes (AVL)
balanced binary trees which permit O(log(N)) read/write operations per
access. The numerical effort associated with building the residue tree
is then proportional to the number of configurations. The number of
matrix elements encoded by the tree, however is proportional to the
expectation value of the square of the lengths of the inner subtrees,
i.e. the inner sum in Eqn.(\ref{eq:mat}). This sum scales with the
number of configurations on a satellite fragment \( (O(N_s^{2}) \) for
the fully selected case. As a result this matrix evaluation scheme 
is very efficient, in tests the expectation values were
computed 200 times faster than with a traditional hash-table based
implementation. This increase in the numerical efficiency permits the
treatment of much larger Hilbert spaces.

Since the different residue trees are independent of one another, we
can implement a relatively simple, scalable parallelization of the
matrix element evaluation on a limited number of nodes, by
distributing the residue trees across the nodes. In our implementation
using 8 nodes of an SGI Power Challenge incurred a total loss of about
12~\% of CPU time compared with a run on a single node. This loss of
efficiency results from the fact that the number of matrix elements
encoded by a given residue tree depends on the number of
configurations containing its residue quantum numbers.

In the \emph{expansion step} for a given reference state \( \left| \Psi
\right\rangle \) we diagonalize
\begin{equation}
\left(
\begin{array}{cc}
\left\langle \Psi |H|\Psi \right\rangle  & 
\left\langle \Psi |H|\phi _{i}\right\rangle \\
\left\langle \phi _{i}|H|\Psi \right\rangle  & 
\left\langle \phi_{i}|H|\phi _{i}\right\rangle  \\
\end{array}
\right)
\label{eq:expand}
\end{equation}
for each trial configuration \( \left| \phi _{i}\right\rangle \) and
use its coefficient for the selection criterion. If the new
configurations \( \left| \phi _{i}\right\rangle \) were mutually
non-interacting this would generate the exact eigenstate of the
system.  The time consuming step in the diagonalization of equation
(\ref{eq:expand}) is the computation of the off-diagonal coupling
matrix element \( \left\langle \phi _{i}|H|\Psi \right\rangle \)
between the reference state and the trial configuration, which can be
accomplished using a similar residue based scheme as for the
evaluation of the matrix elements. Initially we choose a small subset
of configurations with the lowest energies which is diagonalized
exactly. The lowest eigenstate of this Hilbert space is used as the
reference state for the first iteration, for subsequent iterations the
converged state of the previous iteration is used.

For the present system, however, it is not feasible to even consider
all possible trial configurations (N=O($10^{11}$)) in the expansion
loop. In each expansion step we therefore proceed as follows: first we
order the configurations of the last state by the absolute weight of
their coefficients. We then generate the interacting configurations
for the most important configurations until the total weight of the
configurations considered in the reference state exceeds about 90\% of
its norm. {\em Interacting configurations} of a given configuration
are all those that have a nonzero matrix element with the Hilbert
space of the selected configurations. These are typically only a small
fraction (about 1-3\%) of the selected configurations. For these
configurations we apply the selection criteria and gather the selected
subset containing
\( n_{1} \) new configurations. We then generate the interacting
configurations of the next important segment of the reference state
until to total weight of the configurations of the reference state
exceeds 96\% of its norm and generate \( n_{2} \) new configurations
from this set. We continue this process with geometrically decreasing
fraction of the norm until the number of newly generated
configurations \( n_{k} \) is less than 5\% of the total number of
newly generated configurations \( n_{1}+n_{2}+\cdots +n_{k-1}
\). Since configurations with very little weight in the original
wavefunction are unlikely to generate interacting configurations that
will be selected we avoid to even build the full interacting space of
the reference wavefunction.

\begin{figure}[ht]
\vspace*{0.4cm}
{\centering \resizebox*{0.5\columnwidth}{!}
{\includegraphics{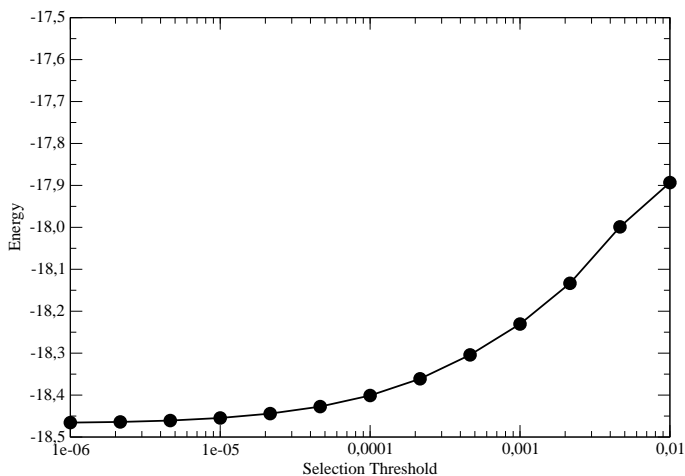}} \par}
\vspace*{0.2cm}
\caption{Convergence of the energy as a function of the selection threshold
\label{conv}
}
\end{figure}

Fig.\ref{conv} illustrates the convergence of the ground state energy
with respect to the selection threshold for the system at hand. It
demonstrates that for selection thresholds less then $1\times 10^{-5}$
the energy is extrapolated with an accuracy of one percent or better.

\vspace*{0.5cm}

Expectation values, such as correlation functions can be easily
computed by adapting the residue driven scheme. One constructs residue
trees corresponding to the appropriate operators and evaluates the
expectation values similar to the energy.

\section{Results}
\label{data}

\subsection{Spin-spin correlations}
\label{corr}

We turn now to the investigation of the ground state magnetic order on
the {\it quantum} s=1/2 antiferromagnetic \Sier \ gasket. The careful
investigation of the spin--spin correlations $ \langle S_i S_j \rangle $
will provide a deeper insight into the ordering behavior.

We begin with the ground state energy (being just the sum over all
nearest-neighbor spin--spin correlations on the \Sier \ gasket). In the
following table we provide the calculated value of the ground state
energy for all finite lattices up to N=42. We provide the site and the
bond average energies (This might help other groups to compare to our
data.)
\begin{table}[ht]
\caption{Number of spins N, site energy $e_s$ and bond energy
$e_b$ for the \Sier \ gaskets up to N=42. (For N=42 the error of
the CSD calculation is given in brackets.)}
\label{ene}
\centerline{
\begin{tabular}{|r||l|l||l|l|}
\hline\noalign{\smallskip}
N   &  Heisenberg & & XY & \\
\noalign{\smallskip}\hline\noalign{\smallskip}
   & $e_s$ & $e_b$ & $e_s$ & $e_b$ \\
6  &    -0.375     & -0.25     &  -0.25     & -0.16666$\bar{6}$ \\
15 &    -0.416125  & -0.231181 &  -0.282024 & -0.156680 \\
42 &    -0.439(5)  & -0.227(9) &  -0.298(8) & -0.154(9) \\
\noalign{\smallskip}\hline
\end{tabular}}
\end{table}
In a quantum spin system with \Ne-like semi-classical ordering the
spin--spin correlation between the spins in one classical sublattice
remains constant for large distances after a slight decay due to the
quantum fluctuations. Therefore we first investigate the
spin--spin correlations between the spins of one classical sublattice
as defined in Sec.\ref{model} (for example all sites with circles
{\Large $\bullet$} in Fig.\ref{fig1}).

The geometrical distance, which is normally the measure on one- or
two-dimensional lattices, is not easily transferable to fractal
objects, such as the \Sier \ gasket. For the \Sier \ gasket we
therefore use the Manhattan distance $r_M$ which counts the minimal
number of steps required to connect one site to the other. Because of
lack of translational symmetry in the \Sier \ gasket we find different
spin--spin correlations between spins having the same Manhattan
distance. In what follows we use a simple averaging procedure over all
spin--spin correlations at a given Manhattan distance.

\begin{figure}[th]
\centerline{
\epsfig{file=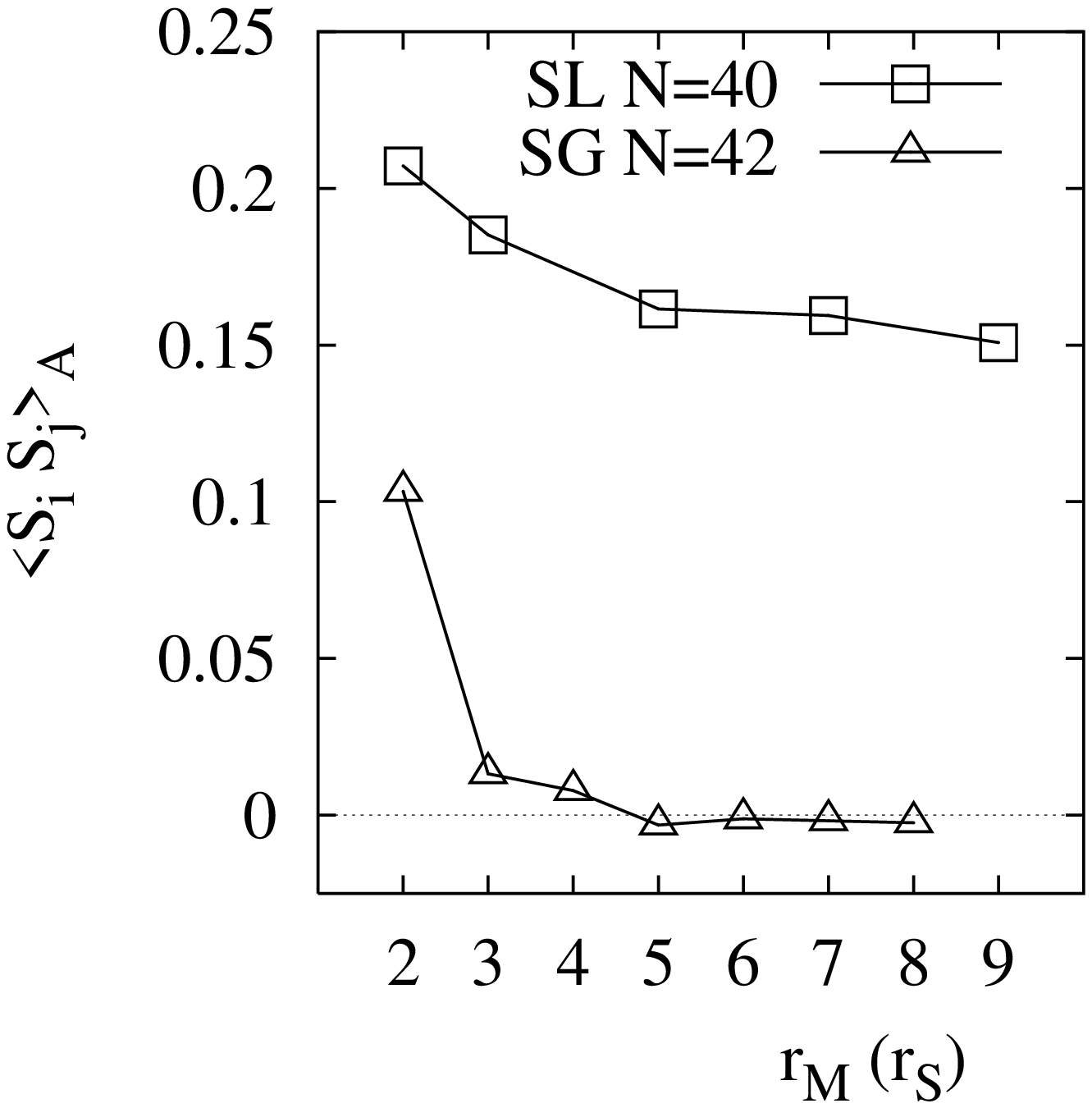,scale=0.45,angle=-90}
\hspace*{1.0cm}
\epsfig{file=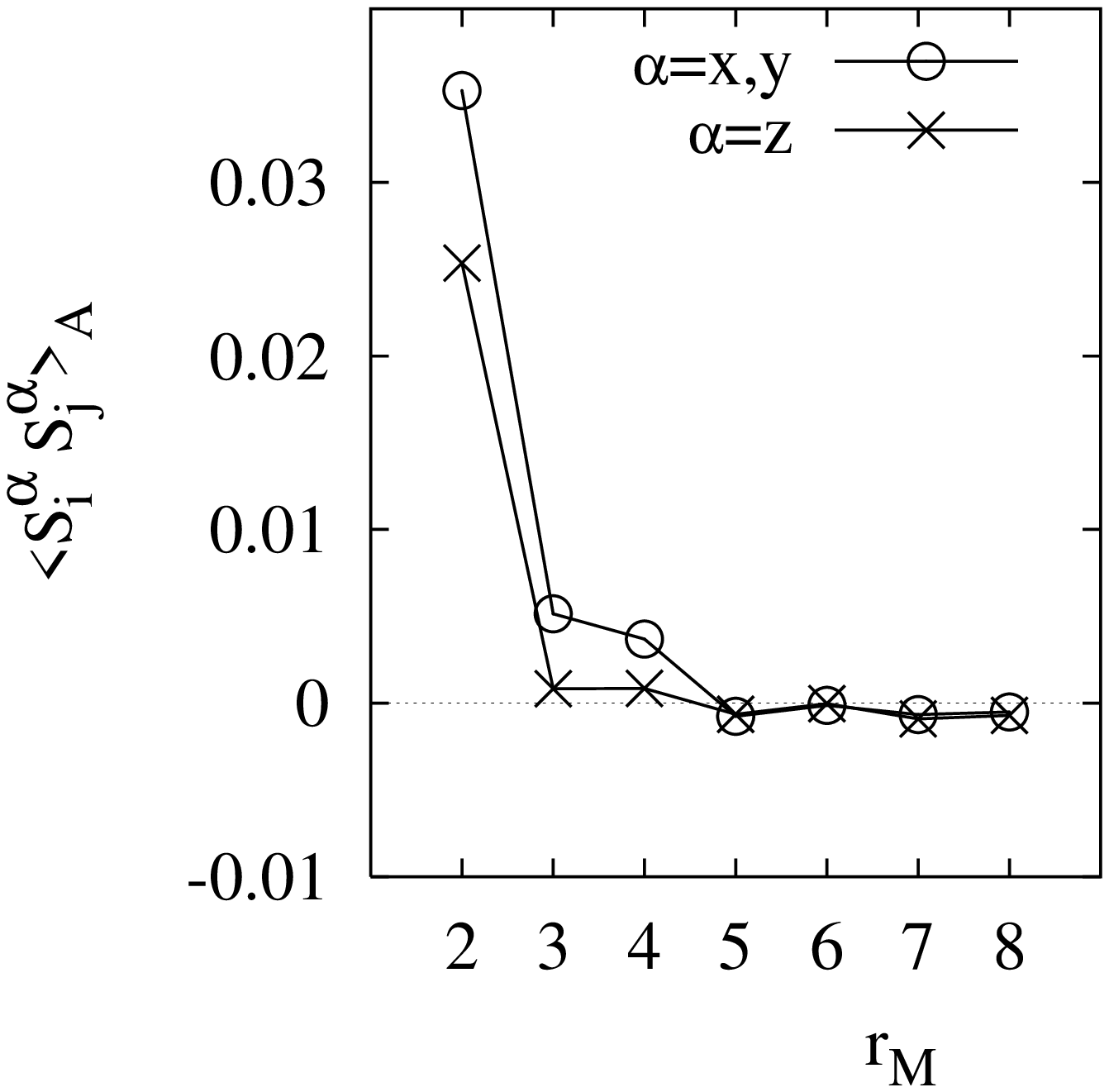,scale=0.45,angle=-90}}
\vspace*{1em}
\caption{The spin--spin correlation between spins on a classical 
sublattice $\langle {S_i S_{j} \rangle }_{i,j \in \bullet}$ 
vs. distance. Left: $\Delta=1$ (Heisenberg model) - 
data for the \Sier \ gasket with N=42 (SG) vs. Manhattan distance $r_M$ 
and corresponding data for a two-dimensional square lattices with N=40
(SL) vs. the shell distance $r_S$; Right: $\Delta=0$ (XY model) - data
for the \Sier \ gasket with N=42 for the x- and z- components of 
$\langle S_i S_j \rangle$.
\label{fig_sisj_r}}
\end{figure}

In our earlier investigations we predicted a disordered ground state
using the exact diagonalization data for N=15 among others. Below we
will compare this data with the new data for N=42 to confirm our
prediction. In Fig.\ref{fig_sisj_r} the spin--spin correlation
$\langle {S_i S_{j} \rangle}_{i,j \in \bullet}$ for the \Sier \
gasket (SG) with N=42 and corresponding data for a two-dimensional
square lattice (SL) with N=40 \cite{schulenburg03,schulenburg03a}
(where magnetic long-range order is well known to exist) is shown.
For the SL data we have chosen the shell distance $r_S$ (where
shell-like circles numbered 1,2,3,$\ldots$ are drawn around a
given site which then connect all neighbors at the given distance).
The \Sier \ gasket shows a dramatic drop in the spin--spin
correlations and for large distances it even changes its sign.
This is in marked contrast to the two-dimensional square lattice
where after a slight decrease due to the quantum fluctuations an
almost constant behavior over distance is observed. This points to
a complete loss of the classical \Ne-like magnetic order in the
quantum Heisenberg AFM on the \Sier \ gasket. Almost the same
behavior can be seen for the XY model with $\Delta=0$. Here we
have to distinguish between the different components of the
spin--spin correlation due to the spin anisotropy: $\langle S^x_i
S^x_j \rangle = \langle S^y_i S^y_j \rangle \ne \langle S^z_i
S^z_j \rangle $. But as one can see the different components of
the spin--spin correlations show similar behavior and the complete
loss of any classical \Ne-like magnetic order.

Even though the classical magnetic order seems to be absent, some other
type of long-range order in the pair correlations might prevail in the
\Sier \ gasket. In order to check this conjecture we calculate the
absolute spin--spin correlation ${ | \langle {S_i S_j \rangle}|}_{i,j
\in N}$ between {\it all} spins (and not just between spins in a
classical sublattices) over $r_M$ (again using an averaging procedure
as described above) and show a semilogarithmic plot of the data.
\begin{figure*}[ht]
\centerline{
\epsfig{file=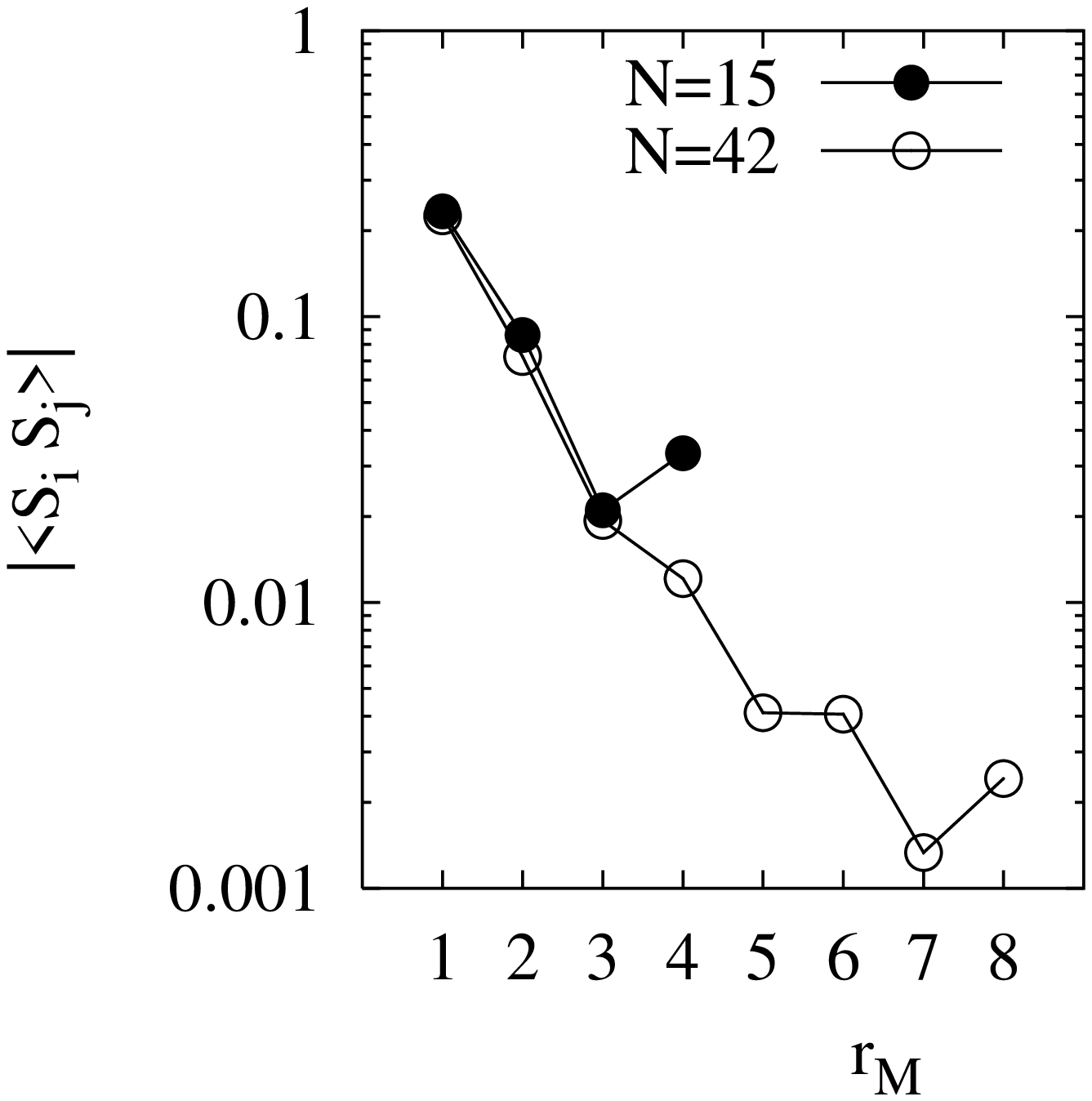,scale=0.450,angle=-90}
\hspace*{1.0cm}
\epsfig{file=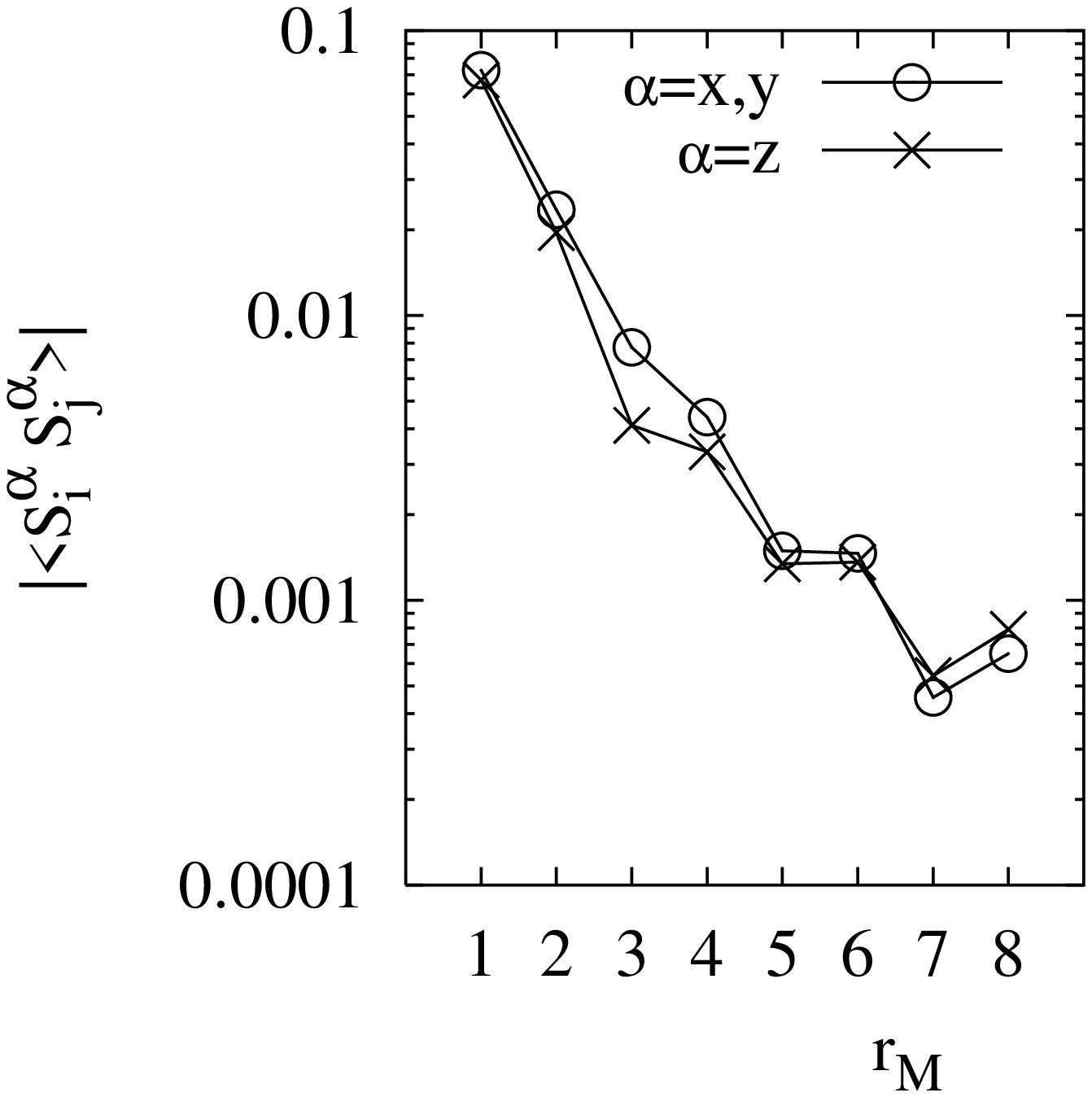,scale=0.450,angle=-90}}
\vspace*{1em}
\caption{The average of all absolute spin--spin correlations
$\overline{| \langle S_i S_j \rangle |}_{i,j \in N} $ vs. 
Manhattan distance $r_M$ of the \Sier \ gasket. 
Left: $\Delta=1$ (Heisenberg model) - 
$\overline{| \langle S_i S_j \rangle |}$ for N=15 and N=42;
Right: $\Delta=0$ (XY model) - 
$\overline{| \langle S_i^x S_j^x \rangle |}$ and 
$\overline{| \langle S_i^z S_j^z \rangle |}$ for N=42.
\label{abs_sisj}}
\end{figure*}
From Fig.\ref{abs_sisj} we deduce that only very short ranged
correlations exist at all in the \Sier \ gasket and this behavior is
independent of the spin anisotropy. For the XY model (right) we
observe again a similar behavior for the different components of
the spin--spin correlation. From the semilogarithmic plot we derive 
a correlation length $\xi \approx 1$, applying ${| \langle
{S_i S_{j} \rangle| }} \sim e^{-(r_M/\xi)}$.  The very small (in
absolute values) upturn of the spin--spin correlations for the largest
separation is presumably a boundary effect of the corner spins of the
\Sier \ gasket. These corner spins have a different coordination
number (only 2 bonds instead of 4) and they are mainly contributing to
the value of this particular spin--spin correlation.

\begin{figure*}[ht]
\centerline{
\epsfig{file=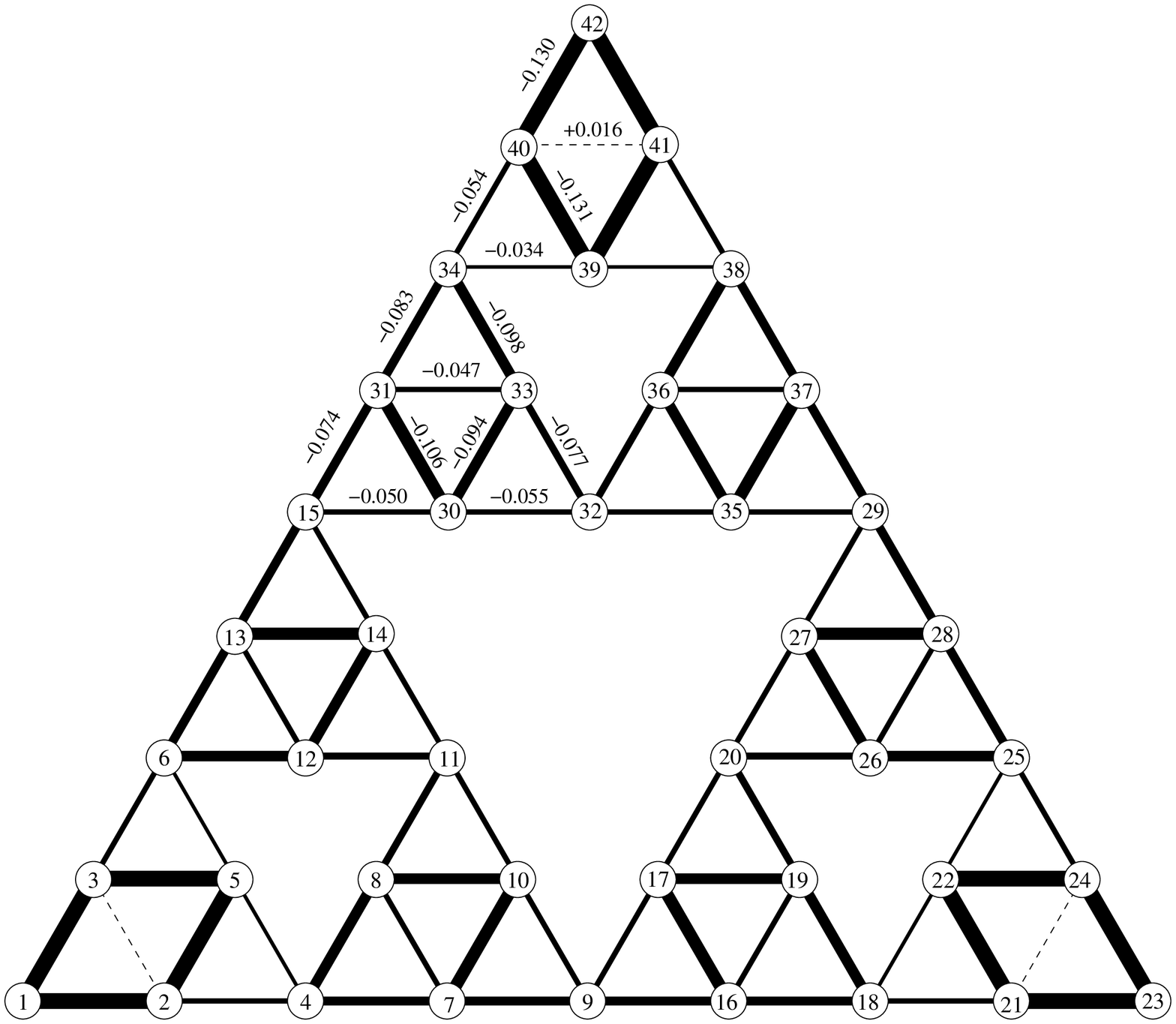,scale=0.35,angle=0}
\epsfig{file=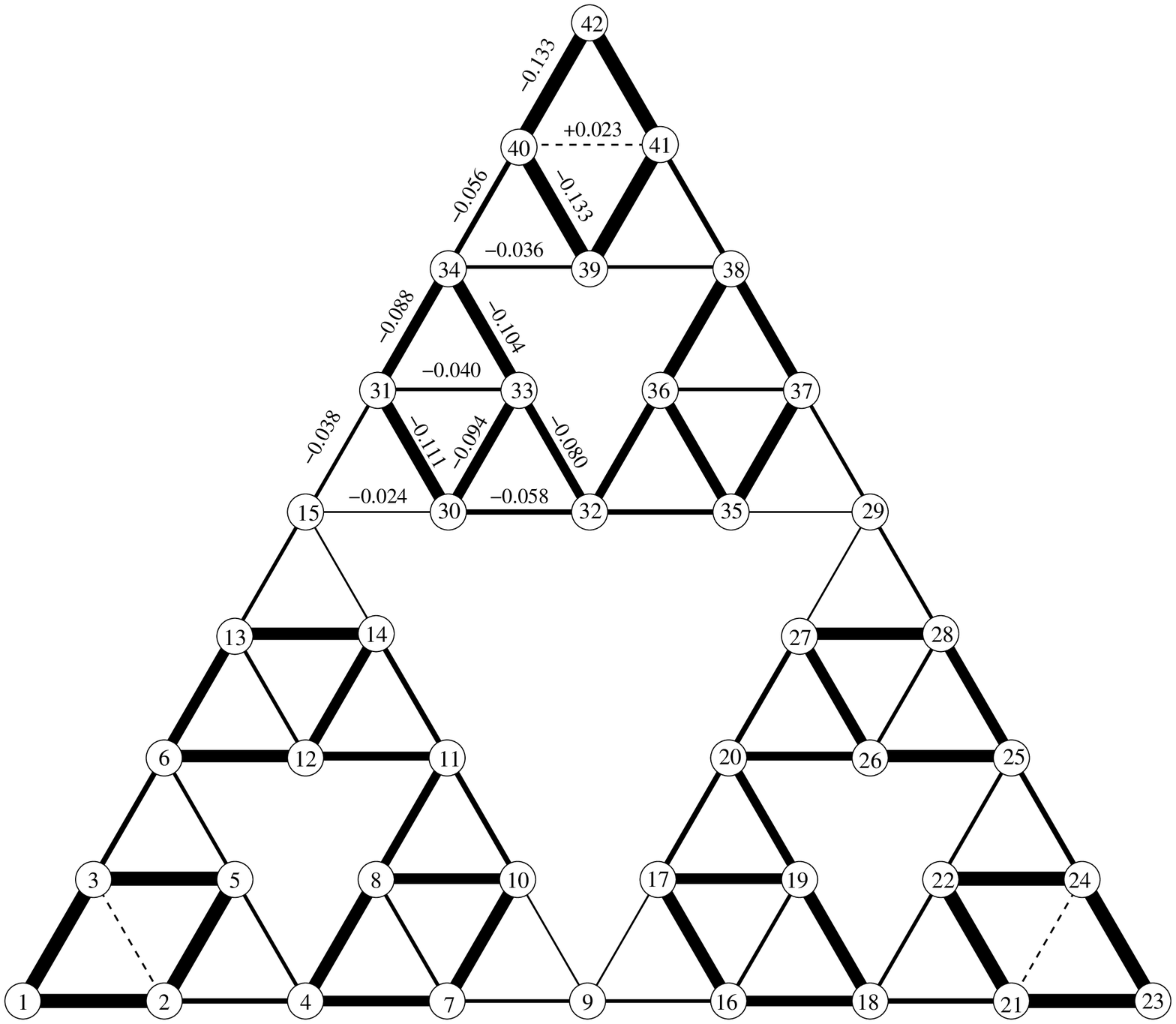,scale=0.35,angle=0}}
\vspace*{1em}
\caption{The nearest neighbor spin--spin correlation on the
\Sier \ gasket with N=42. Left: $\Delta=1$ (Heisenberg model) -
$\langle S^z_i S^z_j \rangle$, Right: $\Delta=0$ (XY model) -
$\langle S^x_i S^x_j \rangle$ (The dotted lines stand 
for a ferromagnetic correlation).
\label{local_corr}}
\end{figure*}

As we observe only short range order in the \Sier \ gasket, we turn to
the investigation of the local order which might exist in this lattice.
We show in Fig.\ref{local_corr} the spin--spin correlations on the
\Sier \ gasket bonds only and chose $\langle S^z_i S^z_j \rangle$
(Heisenberg case) and $\langle S^x_i S^x_j \rangle$ (XY case). The
other components behave quite similar.

We observe in the Heisenberg case (left) as well as in the XY case
(right) a tendency to a plaquette formation at the corner spins. An
example of such a plaquette is seen between the spins 39-40-42-41. Here
the spin--spin correlations are very pronounced between the neighbor
spins on the assumed plaquette (and reach about 80\% of the true
isolated plaquette value as we have checked) and they are apparently
much smaller to the remaining lattice (where for an isolated plaquette
we would observe 0). And even though there is an antiferromagnetic bond
between the spins 40 and 41, the resulting spin--spin correlation
between this two spins is ferromagnetic which points to a beginning
dimer formation between them. The other dimer correlation $\langle
S_{39} S_{42} \rangle$ (value not shown in Fig.\ref{local_corr}) is
0.234 (Heisenberg case) and 0.230 (XY case) and therefore very close to
a true dimer correlation of 1/4. One could argue that another plaquette
may form inside the lattice (one example might be spins 30-31-34-33).
We have found a similar behavior already for the N=15 \Sier \ gasket.
The building of plaquettes in a lattice with strong frustration seems
quite interesting given the fact that in the strong frustrated region
of the $J_1-J_2$ square lattice model one might find a similar behavior
(although this behavior is still under controversial discussion)
\cite{singh99,capriotti00,capriotti01}.

\subsection{Spin gap}
\label{gap}

\begin{figure*}[ht]
\centerline{
\epsfig{file=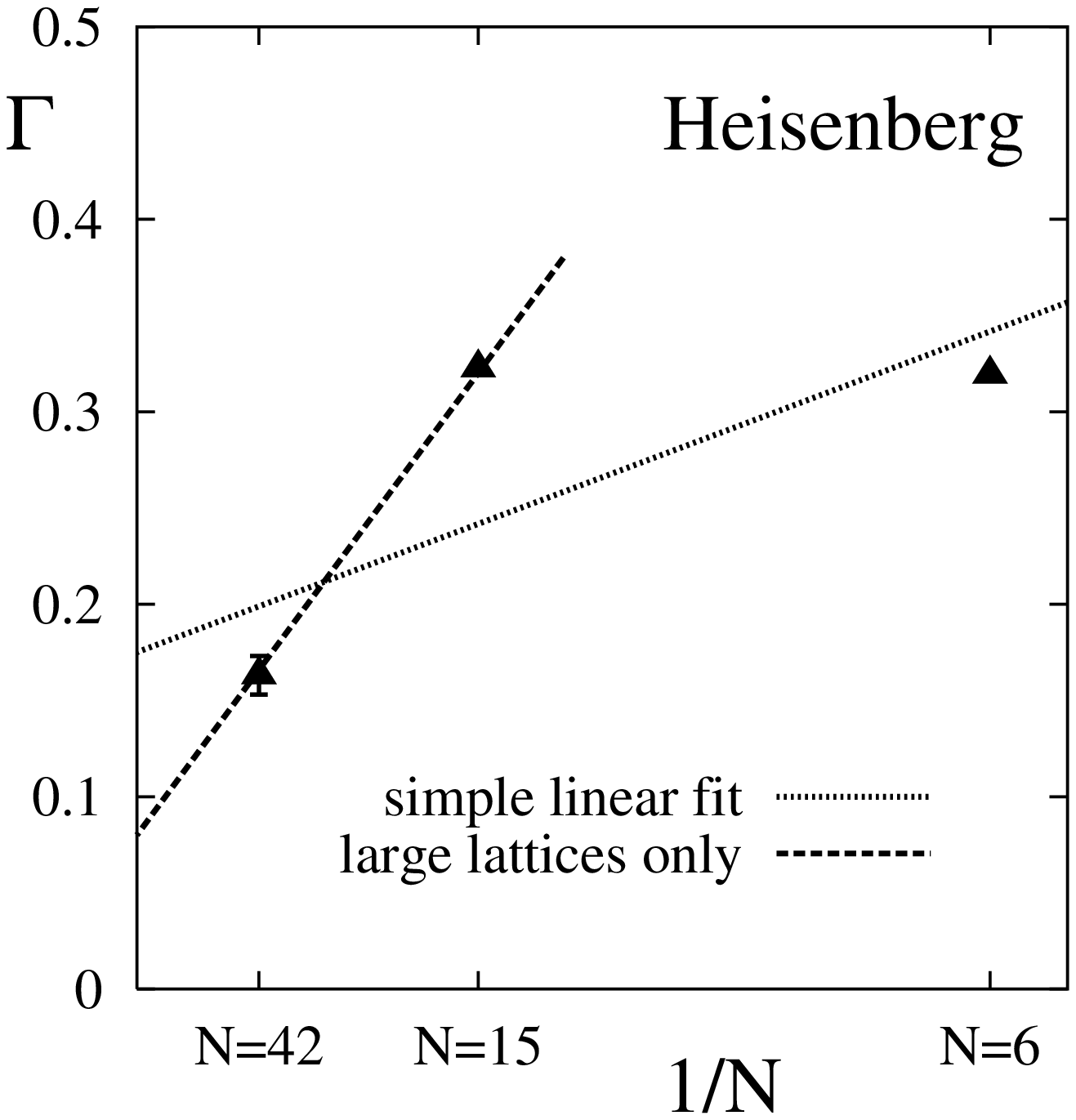,angle=270,scale=0.4}
\hspace*{1.0cm}
\epsfig{file=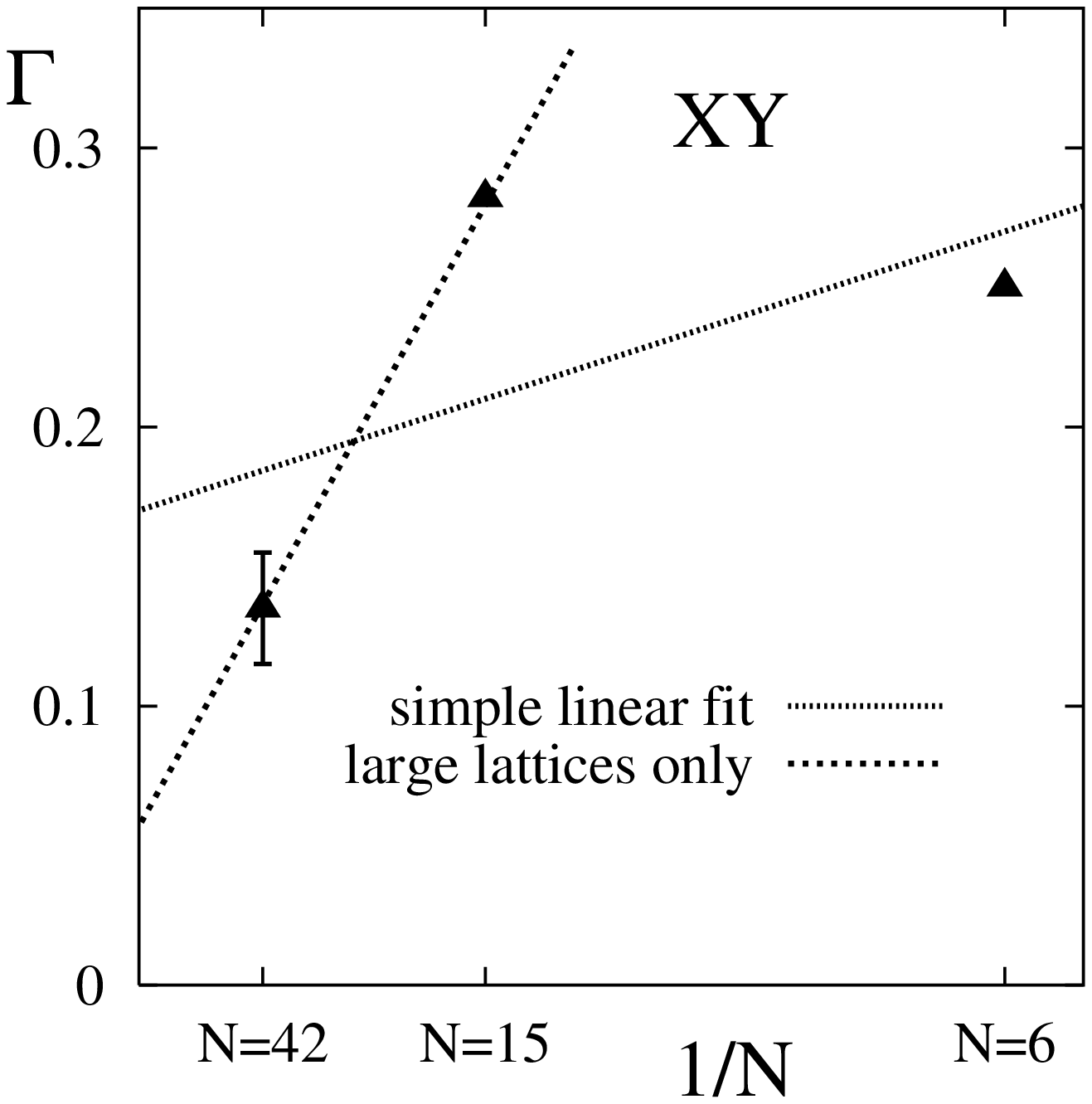,angle=270,scale=0.4}
\vspace*{1em}}
\caption{The spin gap vs. inverse system size for the \Sier \ gasket 
with N=6,15 and 42. Left: $\Delta = 1$ (Heisenberg model); 
Right: $\Delta = 0$ (XY model).
\label{spin_gap}}
\end{figure*}

In this section we investigate the spin gap $\Gamma$, defined as
\be 
\Gamma = E(S^z_{min}+1) - E(S^z_{min}). 
\ee 
The absence of \Ne--like magnetic long-range order is as a rule
accompanied by a finite spin gap and can therefore be used as another
criterion for a disordered ground state \cite{auer94}.  In
Fig.\ref{spin_gap} we present $\Gamma$ for the Heisenberg and XY case.
The data for N=6 and 15 has been presented before but with the CSD we
are able to calculate the spin gap for N=42 as well. The data has to be
analyzed with particular care, because we expect finite size effects
still to be present. If we just make a simple linear fit through the
data we roughly get a value $\Gamma \approx 0.2$. One might also argue
that N=6 is too small to be taken into account for this consideration
and therefore we did another fit with only N=15 and 42. Still we see
that the spin gap $\Gamma$ remains finite, but is reduced to $\Gamma
\approx 0.1$. A similar conclusion holds for the XY model, although the
spin gap appears to be smaller (this behavior has been found in other
investigations too \cite{tomczak01}). Though our finite-size
extrapolation must be taken with particular care we see further
arguments in favor of a finite spin gap and a ground state without
magnetic long--range order.

We mention that the AFM on the \Sier  \ gasket belongs to the
class of frustrated spin systems (like the \kag \ or the
checkerboard lattices) having exactly known localized magnon
eigenstates leading to a macroscopic magnetization jump to
saturation \cite{schulenburg02,richter03}. These localized mag\-nons
can live e.g. on 'hexagons' (large triangles) inside satellite
fragment (e.g. the sites 4,5,6,12,11,8 in Fig.\ref{local_corr}).
For N=42 the magnetization jumps at the saturation field
$h_{sat}=3J$ from $m=17/21$ to saturation $m=1$. Even for N $ \to
\infty$ the height of the jump remains finite since the number of
localized magnons occupying the lattice growths with N.

\subsection{Low temperature thermodynamics}
\label{lowt}

\begin{figure*}[ht]
\centerline{
\epsfig{file=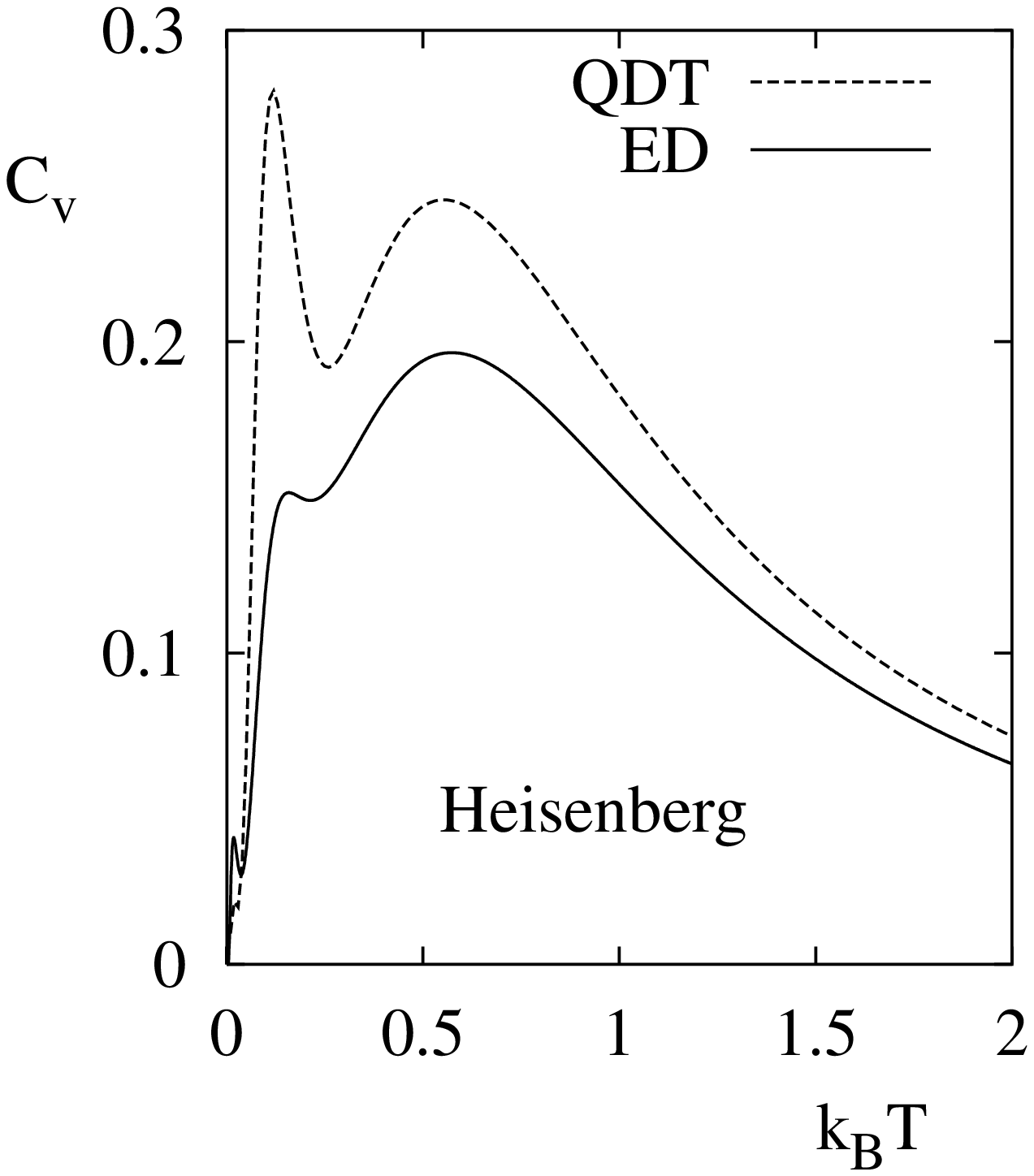,angle=0,scale=0.4}
\hspace{1cm}
\epsfig{file=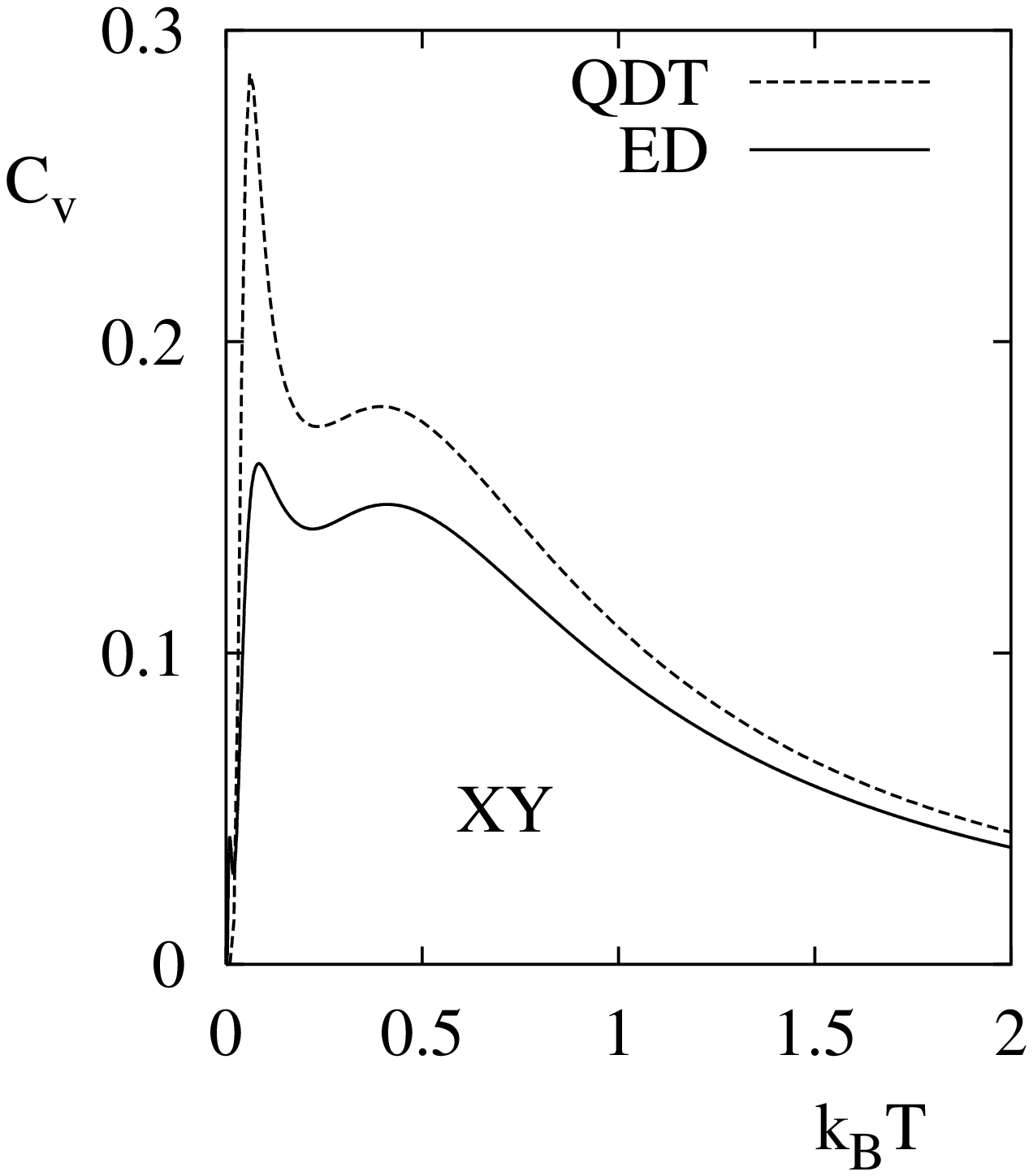,angle=0,scale=0.4}}
\vspace*{0.3cm}
\caption{The low temperature specific heat: full diagonalization data for
the N=15 system and quantum decimation data. Left: $\Delta = 1$
(Heisenberg model); Right: $\Delta = 0$ (XY model).
}.
\label{lowtd}
\end{figure*}

It is known that the low-temperature specific heat is 
closely related to the low-lying excitations of a system
\cite{lhuillier01sep,misguich99,sindzingre00,lhuillier00}. As we have
argued already in Sec.\ref{gap} the low-lying excitations in turn might
show a spin gap behavior and therefore point to a disordered ground
state. We and others have seen in previous investigations a close
connection between an additional low-temperature peak in the specific
heat and a finite spin gap
\cite{voigt98jmmm,sindzingre00,kkubo93,naka96}. Following this
argumentation we will analyze the specific heat $c_v$ of the system
especially in its low-temperature region. We show in Fig.\ref{lowtd} 
$c_v$ for the Hei\-senberg and XY \Sier \ gasket calculated with CD 
and QDT. We note that the $c_v$ results for the Heisenberg case are
identical to those of Ref. \cite{tomczak96prb2} and shown for
comparison purposes.
We observe in both cases additional low-temperature peaks which relate
to two different energy scales relevant in the system. The first energy
scale is connected to the typical broad peak, whereas the second one is
connected to the low-temperature peak and its value is connected to the
finite spin gap $\Gamma_\infty$. This behavior of the specific heat is
typical for all antiferromagnetic systems on "corner sharing triangles"
lattices ({\em kagom\'e} \cite{sindzingre00}, {\em \Sier} 
\cite{tomczak96prb2}, {\em squagome} \cite{tomczak03}). 
In fact, in all those systems the basic unit leading to this behavior
is a $\Delta$-chain \cite{kkubo93,naka96} which {\em shares} spins with
other $\Delta$-chains. ({\em kagom\'e} - 12 spin chain, {\em \Sier} - 6
spin chain, {\em squagome} - 8 spin chain). The RG transformation used
here takes into account the excitation spectrum of 6 spin chain (Fig.
\ref{fig1}, right) and by using Eqn.(\ref {free_en}) one gets a larger
low-temperature peak in comparison to exact diagonalization data.

In both cases an additional low-temperature peak in the specific heat
constitutes an additional argument for a finite spin gap and therefore
for a disordered ground state. The small additional peak in the exact
diagonalization data at even lower temperatures has been attributed to
a finite size effect \cite{voigt01}.

\section{Summary}
\label{summ}

We have carried out a numerical investigation of a s=1/2 quantum
antiferromagnet on the \Sier \ gasket with two types of spin exchange,
Heisenberg and XY. We have used the exact diagonalization, a newly
implemented configuration selective diagonalization approach and a
quantum decimation technique to calculate the spin--spin correlations in
the ground state, the spin gap and the low-temperature specific heat.

The current investigation complements and verifies previous work done
for the \Sier \ gasket 
\cite{tomczak96co,tomczak96prb,tomczak96prb2,voigt98jmmm,voigt01,voigt02}. 
Main progress in the investigation of this fractal many-body system
results from the successful introduction of the configuration selective
diagonalization (CSD). This new method permits the calculation of the
ground state wavefunction (and therefore all correlation functions) and
excited states as well for larger finite quantum spin system. Using CSD
we calculated the spin--spin correlations and the spin gap for N=42. We
note that the method can be applied to other frustrated
low--dimensional quantum spin systems with just moderate changes.

The reported data suggest that the ground state of the \Sier \ gasket
remains disordered for Heisenberg and XY spin exchange. It seems that
the interplay of quantum fluctuations and low dimension prevents any
kind of magnetic long range order in this system. From the available
data we derive a magnetic correlation length $\xi \approx 1$. The
nearest-neighbor spin--spin correlations show a tendency to plaquette
formation. It will be interesting to find out whether or not this
behavior is related to similar findings in the strongly frustrated region
of the $J_1-J_2$ Heisenberg antiferromagnet.

\section{Acknowledgment}

A.V.\ is grateful to the Institute of Molecular Physics of the Polish
Academy of Sciences in Pozna{\'n} (Poland) for kind hospitality during
his stay. This work was supported by the DFG under Grant-Nr. Ri 615/5-1,
KBN Grant-Nr. PO3B 046 14 and by NSF grant ACI-0081789.

\bibliography{all}
\bibliographystyle{unsrt}

\end{document}